\documentclass{emulateapj}
\usepackage{psfig}

\newcommand{\RA}[3]{{#1}^{{\rm h}}{#2}^{{\rm m}}{#3}^{{\rm s}}}
\newcommand{\Dec}[3]{{#1}^{\circ}{#2}'{#3}''}
\newcommand{\E}[1]{\times 10^{#1}}
\newcommand{\twCO}{$^{12}$CO}  \newcommand{\thCO}{$^{13}$CO}

\newcommand{\kpc}{\,{\rm kpc}}
\newcommand{\du}{d_{10.6}}

\begin{document}

\title{Discovery of molecular shells 
associated with supernova remnants. (II) Kesteven~75}

\shorttitle{Molecular shell associated with Kes~75}

\author{
Yang Su\altaffilmark{1,2}, Yang Chen\altaffilmark{1}, Ji
Yang\altaffilmark{2,3}, Bon-Chul Koo\altaffilmark{4}, Xin
Zhou\altaffilmark{1}, Il-Gyo Jeong\altaffilmark{4}, Chun-Guang
Zhang\altaffilmark{1} }

\affil{ $^1$ Department of Astronomy, Nanjing University,
Nanjing 210093, P.R.\ China \\
$^2$ Purple Mountain Observatory, Chinese Academy of Sciences,
Nanjing 210008 \\
$^3$ National Astronomical Observatories, Chinese Academy of
Sciences, Beijing 100012 \\
$^4$ Astronomy program, SEES, Seoul National University, Seoul
151-742, Korea }

\begin{abstract}
The young composite supernova remnant (SNR) Kesteven~75, with a pulsar
wind nebula at its center, has an unusual morphology with a bright
southern half-shell structure
in multiwavelengths. The distance to Kes~75 has long been
uncertain. Aiming to address these issues, we have made millimeter
spectroscopic observations of the molecular gas toward the remnant.
The
$V_{\rm LSR}\sim$ 83--96~km~s$^{-1}$ molecular clouds (MCs) are
found to overlap a large north-western region of the remnant and are
suggested to be located in front of the SNR along the line of sight.
Also in the remnant area, the $V_{\rm LSR}$= 45--58~km~s$^{-1}$ MC
shows a blue-shifted broadening in the \twCO\ (J=1-0) line profile
and a perturbed position-velocity structure near the edge of the
remnant, with the intensity centroid sitting in the northern area of
the remnant. In particular, a cavity surrounded by a molecular shell
is unveiled in the intensity map in the broadened blue wing
(45--51~km~s$^{-1}$), and the southern molecular shell follows the
bright partial SNR shell seen in X-rays, mid-infrared, and radio
continuum. These observational features provide effective evidences
for the association of Kes~75 with the adjacent 54~km~s$^{-1}$ MC.
This association leads to a determination of the kinematic distance
at $\sim10.6\kpc$ to the remnant, which agrees with a location at
the far side of the Sagittarius arm. The morphological coincidence
of the shell seen in multiwavelengths is consistent with a scenario
in which the SNR shock hits a pre-existing dense shell. This dense
molecular shell is suggested to likely represent the debris of the cooled,
clumpy shell of the progenitor's wind bubble proximately behind
the 54~km~s$^{-1}$ cloud.
The discovery of the association with MC provides a possible
explanation for the $\gamma$-ray excess of the remnant.

\end{abstract}


\section{INTRODUCTION}
Massive stars have often not moved far from their matrices in
molecular clouds (MCs) by the time they explode as core-collapse
supernovae because of their short lifetime, and therefore their
remnants are often observed to be associated with MCs. The CO
observations have demonstrated a good positional correlation between
MC complexes and supernova remnants (SNRs) (Huang \& Thaddeus 1986).
About 20 SNRs are known to physically interact with ambient MCs with
convincing evidences (e.g.\ Frail et al.\ 1996). The interaction of
either the progenitors' stellar winds or the SNR shocks with the
MCs can play an important role in the SNRs' evolution and
morphologies. The obscuration by the MCs can also heavily affect the
X-ray morphologies. Some of SNRs have been known to have a
peculiarly asymmetric morphology in multiwavelengths because of the
association with MCs. For example, the western blast wave of SNR
CTB~109 is thought to have been significantly decelerated
by a dense MC, resulting
in an eastern semicircular shape of the remnant (Tatematsu et al.
1987; Wang et al. 1992; Rho \& Petre 1997; Sasaki et al.\ 2004). The
soft X-rays of the northwestern part of SNR~3C~391 are substantially
extincted by the MC in which the SNR is embedded (Rho \& Peter 1996;
Wilner et al.\ 1998; Chen \& Slane 2001; Chen et al.\ 2004).
Following this notion, other SNRs with highly asymmetric
(especially, half-brightened) morphologies, such as Kesteven~75 and
Kesteven~69 (Zhou et al.\ 2008; hereafter Paper~I),
are thus of particular interests.

Kesteven~75 (G29.7$-$0.3) is one of the composite SNRs, exhibiting
both the Crab-like and shell-type properties (Becker \& Helfand
1984; Blanton \& Helfand 1996). A young energetic X-ray pulsar, PSR
J1846-0258, discovered using the {\sl Rossi X-ray Timing Explorer
(RXTE)}, was localized to within an arcminute of the remnant center
using an {\sl ASCA} observation (Gotthelf et al. 2000). The bright
point-like source at the core of the remnant was later resolved with
the high resolution {\sl Chandra} observation (Helfand et al.\
2003). Notably, both the Very Large Array (VLA) 1.4~GHz radio
and {\sl Spitzer} 24$\mu$m mid-infrared (IR)
observations show a semicircular partial shell (3.5$\arcmin$ in
diameter) in the south (Becker \& Helfand 1984; Morton et al.\
2007). The {\sl Chandra} X-ray image of Kes~75 clearly exhibits a
synchrotron-emitting pulsar wind nebula (PWN) in the center, two
bright elongated patches (in the southeast and southwest,
respectively) along the radio and IR partial shell, and very faint
diffuse emission in the northern part (Helfand et al.\ 2003).
Recently, by deep {\sl Chandra} X-ray observations, the X-ray
softening and temporal flux enhancement of the central pulsar were
revealed (Kumar \& Safi-Harb 2008), and a detailed spatial and
spectral analysis of the PWN was made (Ng et al.\ 2008); in both
works, a distance 6~kpc was used.

On the other hand, the distance to Kes~75 is under controversy. By
HI absorption observations, Caswell et al.\ (1975) gave a range
$\sim6.6$--19$\kpc$ for the kinematic distance, covering the value
$\sim19\kpc$ derived by Milne (1979) from the $\Sigma$--$D$
relationship, while Becker \& Helfand (1984) suggested
$\sim21$~kpc. With a higher-sensitivity HI observation,
Leahy \& Tian (2008, hereafter LT08) provided a small estimate of
$\sim5.1$--7.5~kpc. McBride et al.\ (2008) favor a distance as
small as $\sim3\kpc$, so as to reconcile the problem of the
otherwise very large efficiency in converting the central pulsar's
spin-down power into X- and $\gamma$-ray luminosity.

Investigating the physical relation between Kes~75 and its ambient
MCs may be very helpful to understanding its evolution and
irregular morphology as well as clarifying its distance. In this paper,
we present our millimeter CO observations toward this remnant.
In \S~2, we describe the observations and the data reduction. The
main observational results are presented in \S~3,
and discussion and conclusions
are in \S~4 and \S~5, respectively.

\section{OBSERVATIONS AND DATA REDUCTION}
The observation was first made in the $^{12}$CO~($J$=1-0) line (at
115.271~GHz) in 2006 April using
the 6-meter millimeter-wavelength telescope of the Seoul Radio
Astronomy Observatory (SRAO) with single-side band filter and the
1024-channel autocorrelator with 50-MHz bandwidth.
The half-power beam size at 115~GHz is $2'$
and the typical rms noise level was about 0.2~K at the
0.5~km~s$^{-1}$ velocity resolution. The radiation temperature is
determined by $T_{\rm R}=T_{\rm A}/(f_b\times\eta_{\rm mb})$, where
$T_A$ is the antenna temperature, $f_b$ the beam filling factor (assuming
$f_b\sim$~1), and $\eta_{\rm mb}$ the main beam efficiency
($\sim$~75$\%$). We mapped the $10' \times 10'$ area covering Kes~75
centered at $(\RA{18}{46}{25}.0,\Dec{-02}{59}{00})$ with grid
spacing $\sim 1'$.

The follow-up observations were made in the $^{12}$CO~($J$=1-0) line
(at 115.271~GHz), $^{13}$CO~($J$=1-0) line (110.201~GHz), and
C$^{18}$O~($J$=1-0) line (109.782~GHz) during 2006 November-December
and 2007 October-November using the 13.7-meter millimeter-wavelength
telescope of the Purple Mountain Observatory at Delingha (hereafter
PMOD). The spectrometer has 1024 channels, with a total bandwidth of
145~MHz (43~MHz) and the velocity resolution of 0.5~km~s$^{-1}$
(0.2~km~s$^{-1}$) for $^{12}$CO ($^{13}$CO and C$^{18}$O).
The half-power beamwidth of the telescope is 54$\arcsec$ and the
main beam efficiency $\eta_{\rm mb}$ is 67$\%$. We mapped the
source centered at $(\RA{18}{46}{25}.0,\Dec{-02}{59}{18})$ with
grid spacing $0.5\arcmin$ in the inner ($4.5'\times5'$) region
(rms noise $<0.1$~K) and $1\arcmin$ in the outer region (out to
$10'\times10'$, rms noise $\sim0.3$~K). All the CO observation
data were reduced using the GILDAS/CLASS package developed by
IRAM\footnote{http://www.iram.fr/IRAMFR/GILDAS}.

The {\sl Chandra} X-ray and {\sl Spitzer} 24~$\mu$m mid-IR
observations are also used. We revisited the {\sl Chandra} ACIS
observational data of SNR Kes~75 [ObsIDs 748 (PI: D.J.\ Helfand),
6686, 7337, 7338, \& 7339 (PI: P.\ Slane) with total exposure time
of 196~ks]. We reprocessed the event files (from Level 1 to Level 2)
using the CIAO3.4 data processing software to remove pixel
randomization and to correct for CCD charge-transfer inefficiencies.
The overall light curve was examined for possible contamination from
a time-variable background. The reduced data, with a net exposure
of 183~ks, were used for our analysis. The mid-IR 24~$\mu$m
observation used here was carried out as the program of the
Breaking Down the Spectra of Pulsar Wind Nebulae (PID: 3647, PI: P.\
Slane) with the Multiband Imaging Photometer (MIPS) (Rieke et al.
2004). The Post Basic Calibrated Data (PBCD) of the 24~$\mu$m mid-IR
are obtained from {\sl Spitzer} archive. The 1420~MHz continuum and
HI-line data come from the archival VLA Galactic Plane
Survey (VGPS) (Stil et al. 2006).

\section{RESULTS}
\subsection{Velocity Structure of Molecular Gas Components}

\begin{figure}[tbh!]
\centerline{ {\hfil\hfil
\psfig{figure=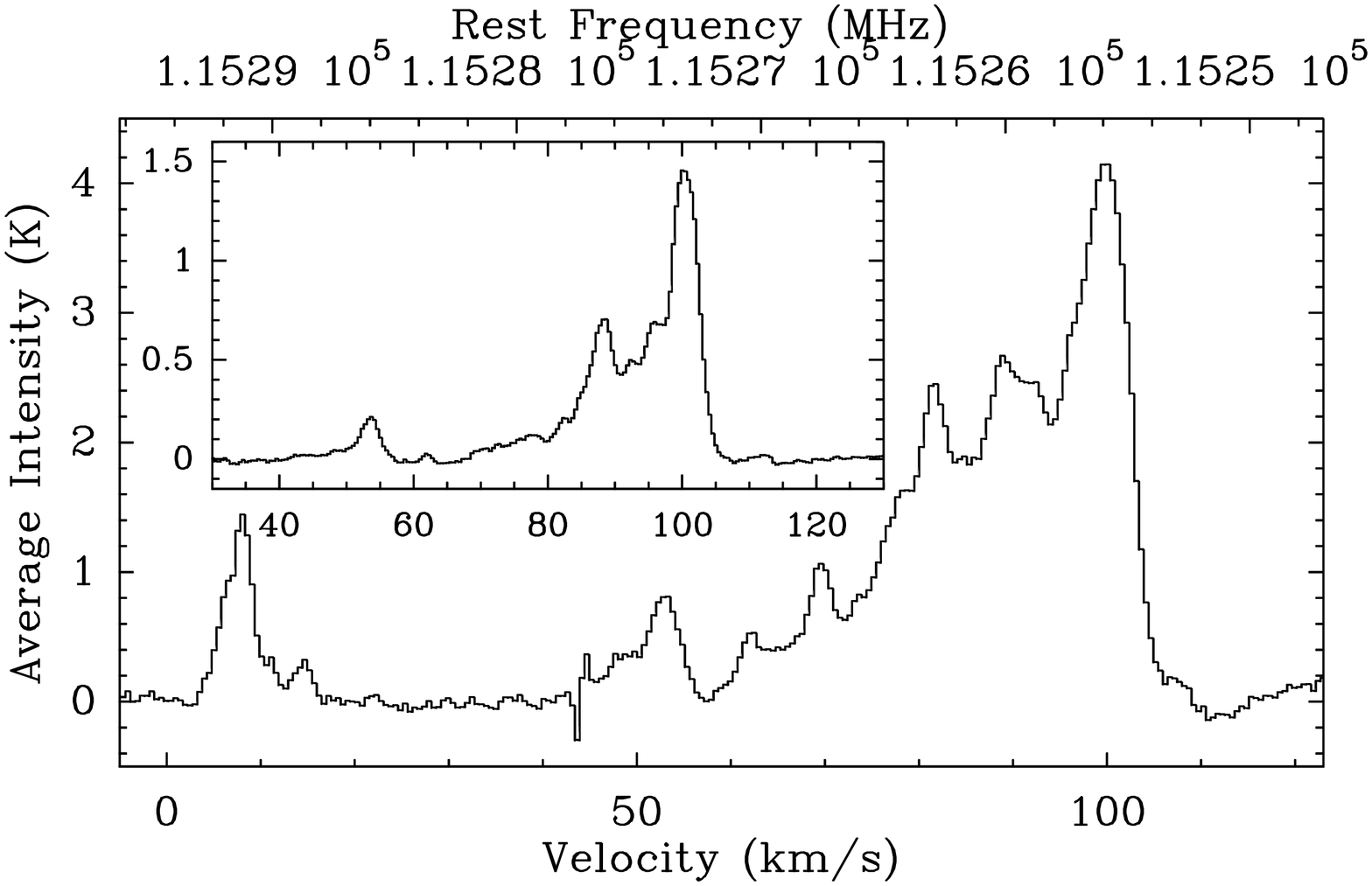,height=2.2in,angle=270, clip=} \hfil\hfil}}
\caption{$^{12}$CO($J$=1-0) spectrum from the SRAO observation
($0\leq V \leq120$ km~s$^{-1}$) toward SNR Kes~75 in a
$10'\times10'$ area centered at
$(\RA{18}{46}{25}.0,\Dec{-02}{59}{00})$. In the little box is the
\thCO($J$=1-0) spectrum from the PMOD observation ($20\leq V
\leq120$ km~s$^{-1}$) toward SNR Kes~75 in a $4.5'\times5'$ area
centered at $(\RA{18}{46}{25}.0,\Dec{-02}{59}{18})$.}
\label{f:srao_sp}
\end{figure}

We show in Figure~\ref{f:srao_sp} the $^{12}$CO~($J$=1-0) spectrum
from the SRAO observation in the velocity range 0--120~km~s$^{-1}$
toward Kes~75 in a $10'\times10'$ area.
Beyond this range, virtually no $^{12}$CO emission is detected at
$-$110~km~s$^{-1}$$\leq$ $V_{\rm LSR}\leq$ 0~km~s$^{-1}$ and
$V_{\rm LSR}\geq$ 115~km~s$^{-1}$ in the PMOD observation. The CO
emission in the direction of SNR~Kes~75 is present in a broad velocity
range between 3 and 110~km~s$^{-1}$, characterized by several
prominent peaks.
The first \twCO\ component is represented by the primary emission
peak in the interval $V_{\rm LSR}$=3--17~km~s$^{-1}$. We call the
molecular gas at 40--60~km~s$^{-1}$ as the second component and
will show its association with the SNR. The other \twCO\
emission peaks in the interval 60--110~km~s$^{-1}$ are referred to
as the third component of MC complex, although the peaks in this
interval are not necessarily related to each other. The first and
the third velocity components are probably corresponding to the
Aquila Rift (Dame et al.\ 2001) and the MCs in the Scutum and
4-kpc arms (Dame et al.\ 1986), respectively. The second velocity
component displays a narrow Gaussian profile interval of
49--58~km~s$^{-1}$ around $\sim$54~km~s$^{-1}$ and an additional
broad line emission structure interval of 45--51~km~s$^{-1}$ (with
an average 6~km~s$^{-1}$ blueshift). In the \thCO~($J$=1-0)
observation ($20\leq V \leq120$ km~s$^{-1}$), there are the
counterparts of the second and third molecular
components. No signal is detected in the range 20--120~km~s$^{-1}$
in the PMOD C$^{18}$O~($J$=1-0) observation.

\begin{figure*}[tbh!]
\centerline{ {\hfil\hfil
\psfig{figure=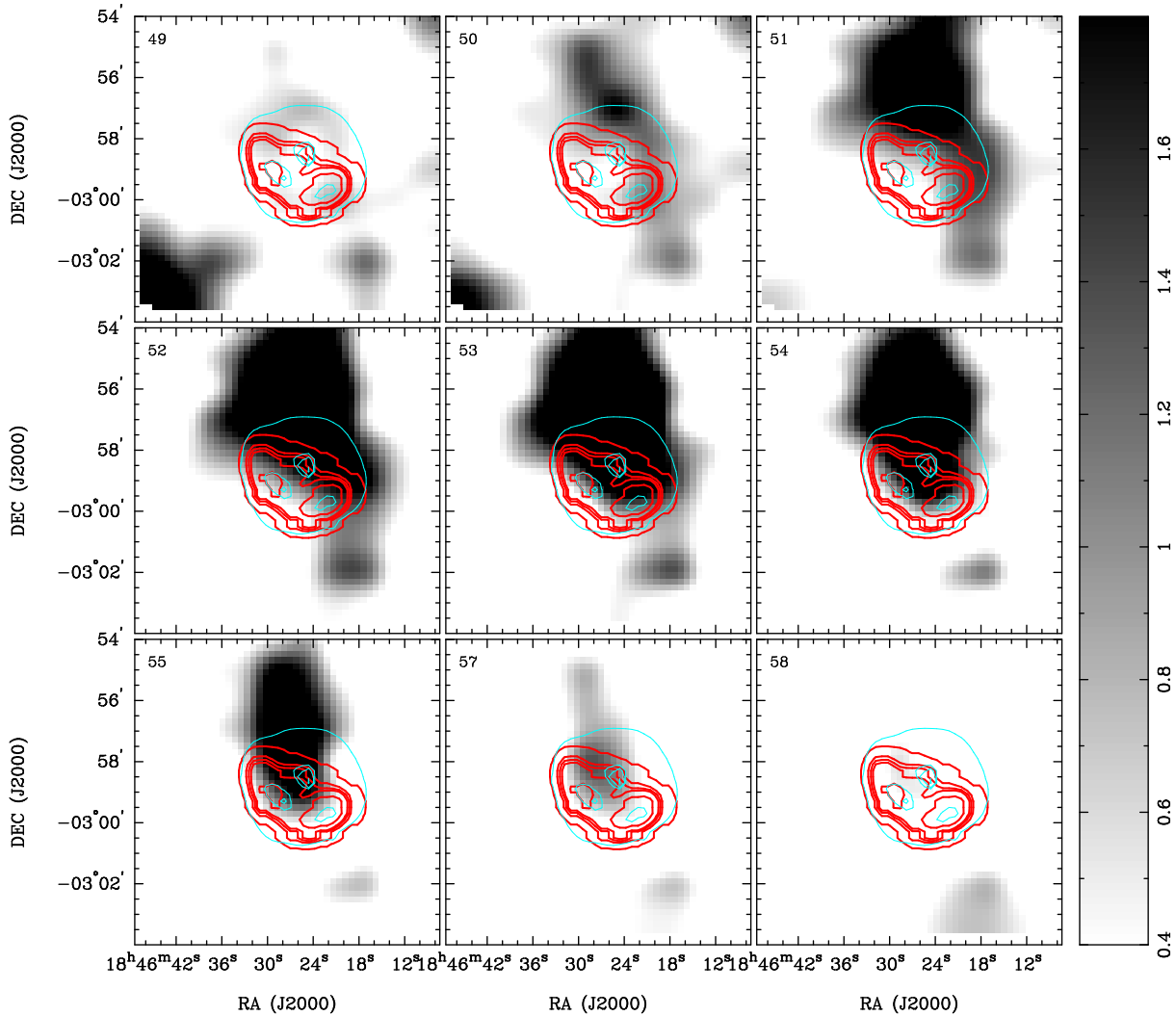,height=5.5in,angle=270, clip=} \hfil\hfil}}
\caption{The SRAO $^{12}$CO ($J$=1-0) intensity channel maps
between 49 km~s$^{-1}$ and 58 km~s$^{-1}$ (smoothed to a
resolution of $0.2'$ by interpolation), overlaid with the 1.0--7.0~keV
X-ray ({\em cyan, thin}) and 1.4~GHz radio ({\em red, thick}) continuum emission
contours (from the NRAO VLA Sky Survey).} \label{f:srao_im1}
\end{figure*}

\begin{figure*}[tbh!]
\centerline{ {\hfil\hfil
\psfig{figure=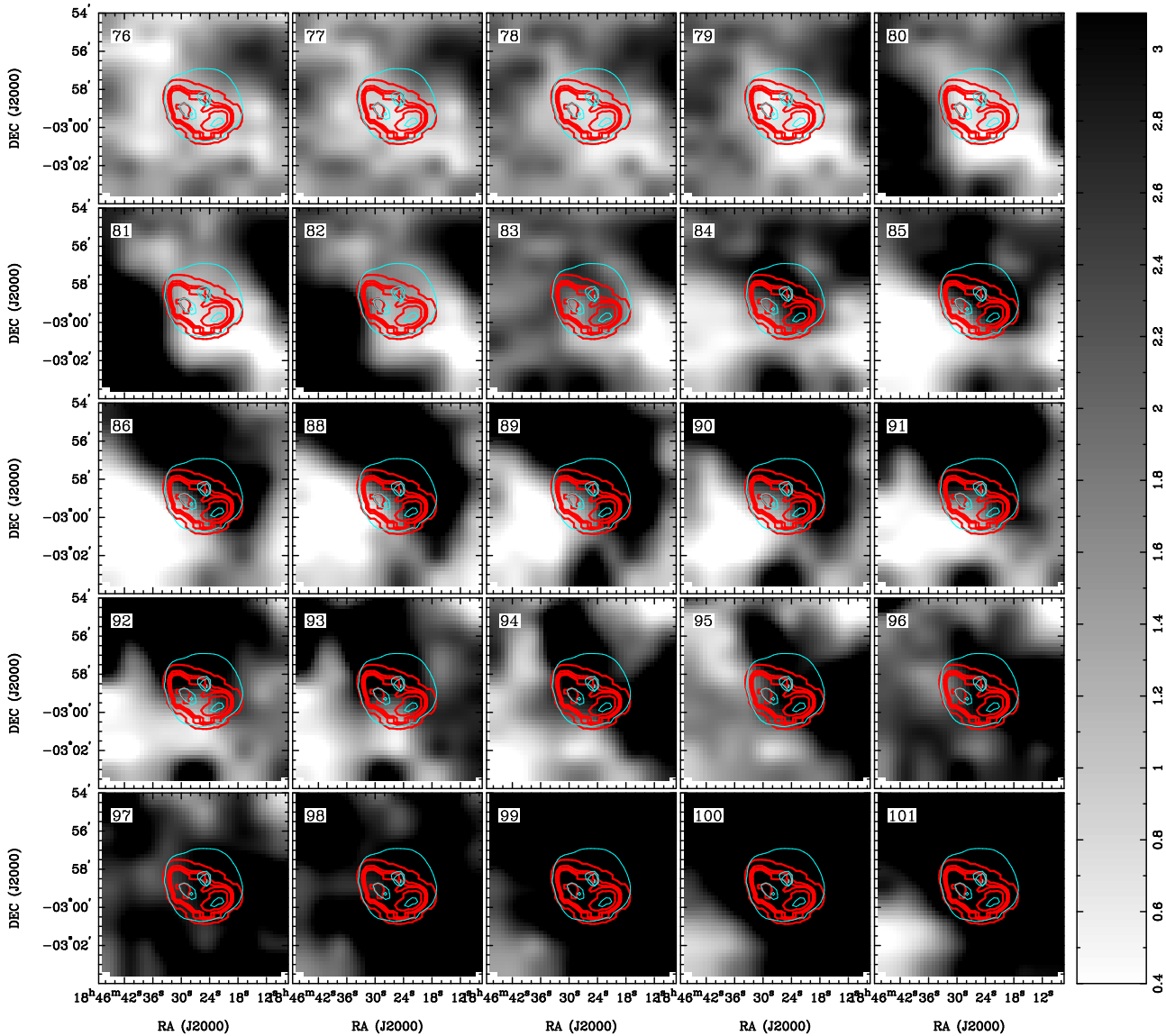,height=6.0in,angle=270, clip=} \hfil\hfil}}
\caption{The SRAO $^{12}$CO($J$=1-0) intensity channel maps
between 76~km~s$^{-1}$ and 101~km~s$^{-1}$ (smoothed to a
resolution of $0.2'$ by interpolation), overlaid with the 1.0--7.0~keV
X-ray ({\em cyan, thin}) and 1.4~GHz radio ({\em red, thick}) continuum
emission contours (from the NRAO VLA Sky Survey).}
\label{f:srao_im2}
\end{figure*}

We made the $^{12}$CO-emission intensity channel maps with
$\sim1$~km~s$^{-1}$ velocity intervals. No evidence is found for
the positional correlation between the first CO component and SNR
Kes~75. However, the second and third components seem to have
positional correlations with the remnant. Figures~\ref{f:srao_im1}
and ~\ref{f:srao_im2} show the $^{12}$CO channel maps in the
intervals 49--58 km~s$^{-1}$ and 76--101 km~s$^{-1}$ (the main
bodies of the second and third velocity components), respectively,
with 1$\arcmin$ grid pointing. The channel maps are overlaid with
the contours of the {\sl Chandra} 1.0--7.0~keV X-ray and VLA
1.4~GHz radio continuum emission, respectively. In morphology,
both the X-ray and radio emissions show bright half in
the south and faint half in the north. It is interesting that the
49--58~km~s$^{-1}$ cloud (as a part of the second component) is
projectionally situated north to the radio- and X-ray-bright
southern half of the SNR, seeming to cover the faint northern
half. The CO emission around 51~km~s$^{-1}$ also appears to
surround the SNR from the north to the west. As parts of the third
component of MC complex, the $V_{\rm LSR}\sim$ 83--96~km~s$^{-1}$
MCs overlap a large north-western region of Kes~75, and the
$\sim97$--101~km~s$^{-1}$ MCs completely overlap the remnant
region (Fig.~\ref{f:srao_im2}). The brightness anti-correlation
between the two $^{12}$CO line components and the X-ray and radio
(as well as mid-IR 24$\mu$m, \S\ref{54_MC}) emissions is clear in
a large scale.

\subsection{The $V_{\rm LSR}\sim54$~km~s$^{-1}$ Molecular Cloud Component }
\label{54_MC}

\begin{deluxetable}{cccc}
\tablecaption{Parameters of the Gaussian profile at $V_{\rm
LSR}$=49-58~km~s$^{-1}$ derived from the PMOD observation.}
\tablehead{\colhead{Line}
          &\colhead{Central velocity}
          &\colhead{Width of line\tablenotemark{a}}
          &\colhead{Mean Intensity\tablenotemark{b}} \\
\colhead{} & \colhead{(km~s$^{-1}$ ) } & \colhead{(km~s$^{-1}$ ) } &
\colhead{(K)} } \startdata
$^{12}$CO & 53.7  & 3.7 & 3.16 \\
$^{13}$CO & 53.6  & 3.3 & 0.44 \\
\enddata
\tablenotetext{a}{The half velocity width of the line.}
\tablenotetext{b}{The region of the MC in the interval
49-58~km~s$^{-1}$ is about 27~arcmin$^2$.}
\end{deluxetable}

In order to obtain detailed clues to the possible association of the
cloud components with the SNR, we have analyzed the second and third
velocity components in the inner $4.5'\times5'$ area [centered at
($\RA{18}{46}{25}.0,\Dec{-02}{59}{18}$)] based on the PMOD
observation.

\begin{deluxetable*}{cclll}
\tablecaption{Parameters for the MC derived from the PMOD
observation.} \tablehead{ \colhead{Region} &\colhead{$V_{\rm LSR}$
interval} &\colhead{H$_2$ column density\tablenotemark{a}}
          &\colhead{Virial mass\tablenotemark{b}}
          &\colhead{Molecular mass}\\
          \colhead{} & \colhead{(km~s$^{-1}$ ) } &
           \colhead{$N$(H$_2$)~(10$^{20}$cm$^{-2}$) } & \colhead{$M_{\rm vir}$
($10^3M_{\odot}$) } & \colhead{$M$ ($10^3M_{\odot}$)} } \startdata
Main body & 49--58 & 22.4 / 21.5 & $26.6\du$\tablenotemark{c} &
$12.6\du^2$ / $12.1\du^2$\tablenotemark{c}\\ \hline
 Southern region\tablenotemark{d} & 45--58 & 14.8 &  &
$2.7\du^2$\tablenotemark{c}\\
 \multicolumn{2}{c}{Gaussian around 54~km~s$^{-1}$} & 9.6 &  &
$1.8\du^2$ \tablenotemark{c}\\
 \multicolumn{2}{c}{Gaussian subtracted} & 5.2 &  &
$0.9\du^2$\tablenotemark{c}
\enddata
\tablenotetext{a}{See text for the two methods used for the column
density derivation.}\tablenotetext{b}{The virial mass is obtained
using the formula $M_{\rm vir}=210(r/\mbox{1pc})(\Delta
v/\mbox{1km~s}^{-1})^2$ $M_{\odot}$, where $\Delta v$ is the FWHM of
the line (Caselli et al.\ 2002).} \tablenotetext{c}{Parameter $\du$
is the distance to the cloud in units of 10.6~kpc
(\S\ref{distance}).} \tablenotetext{d}{A sky region about
9~arcmin$^2$, see Fig.\ref{f:close-up}.}
\end{deluxetable*}

The observed parameters of the CO lines and the derived column
density and cloud mass for the second velocity component
(in interval 49--58~km~s$^{-1}$) are summarized in Table~1 and
Table~2, respectively. Two methods have been used in the
derivation of the molecular column density and similar results
are yielded. In the first method, the H$_2$ column density is
calculated by adopting the mean CO-to-H$_2$ mass conversion factor
$1.8\E{20}$~cm$^{-2}$K$^{-1}$km$^{-1}$s (Dame et al.\ 2001;
similar values also derived by Strong \& Mattox 1996 and Hunter et
al.\ 1997). In the second method, on the assumption of local
thermodynamic equilibrium (LTE) and $^{12}$CO ($J$=1-0) line being
optically thick, the $^{13}$CO column density is converted to the
H$_2$ column density using $N$(H$_2$)/$N(^{13}$CO) $\approx7\E{5}$
(Frerking et al.\ 1982). In the estimate of the LTE mass of the
molecular gas, a mean molecular weight per H$_2$ molecule
$2.76m_{\rm H}$ has been adopted. Assuming the line of sight (LOS)
size of the cloud similar to its apparent size $\sim4'$, we
estimate the molecular density as $n({\rm
H}_2)\sim60\du^{-1}\,{\rm cm}^{-3}$, where $\du=d/(10.6$~kpc) is
the distance to the MC/SNR in units of the reference value (see
\S\ref{distance} for the distance determination).

\begin{figure*}[tbh!]
\centerline{ {\hfil\hfil
\psfig{figure=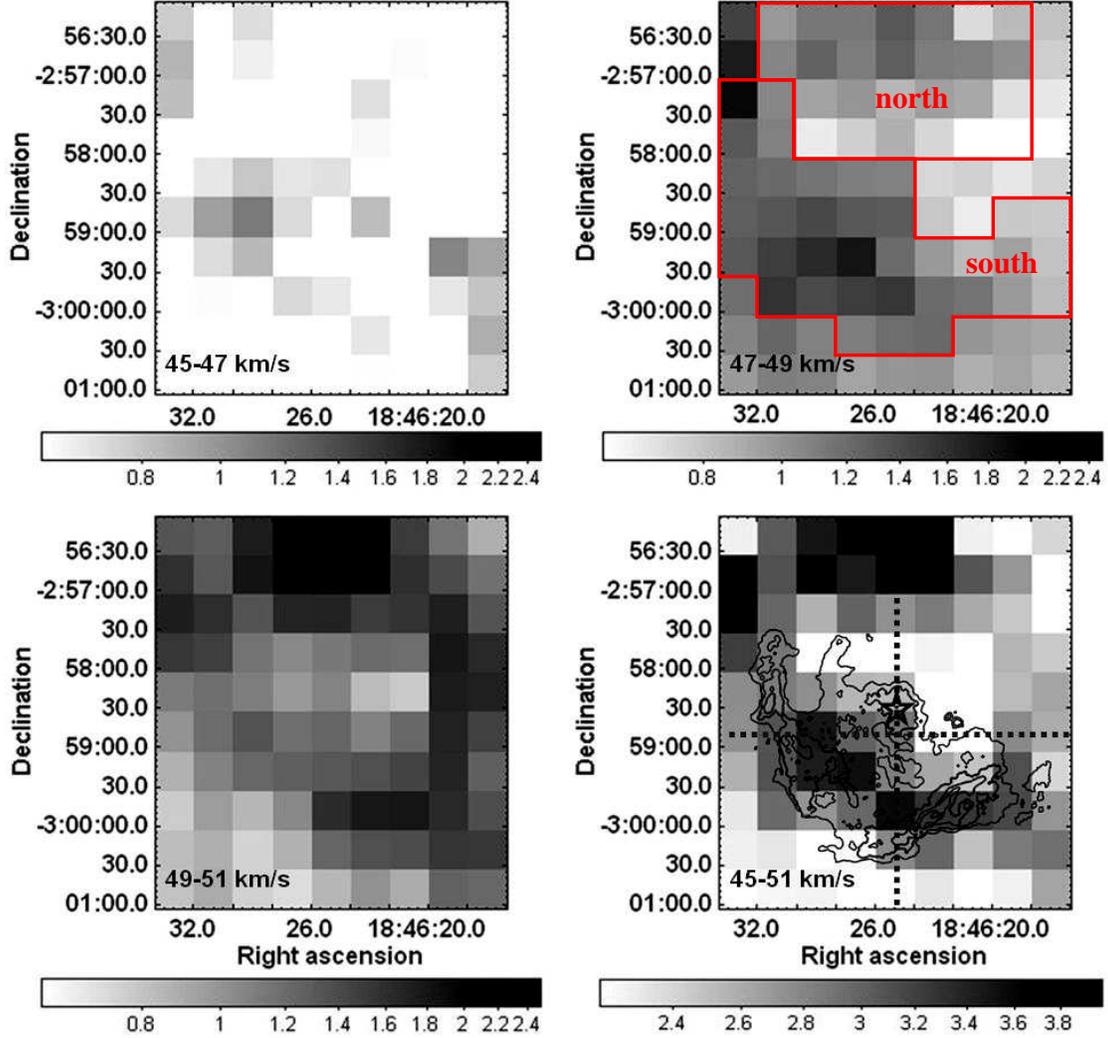,height=5.5in,angle=0, clip=} \hfil\hfil}}
\caption{The PMOD $^{12}$CO (J=1-0) intensity maps
in the intervals 45--47, 47--49, 49--51, \& 45--51~km~s$^{-1}$. The
two regions (6.5~arcmin$^2$ {\em upper} and 9~arcmin$^2$ {\em
lower}) in the {\em upper-right} panel are defined for the
CO-spectrum extraction. In the {\em lower-right} panel, the \twCO\
intensity map is overlaid with the 6~cm radio continuum contours
(Becker \& Helfand 1984), the star stands for the position of the
pulsar, and the dotted lines indicate the E-W and N-S cuts for the
PV diagrams (Fig.~\ref{f:pv}).} \label{f:close-up}
\end{figure*}

\begin{figure}[tbh!]
\centerline{ {\hfil\hfil
\psfig{figure=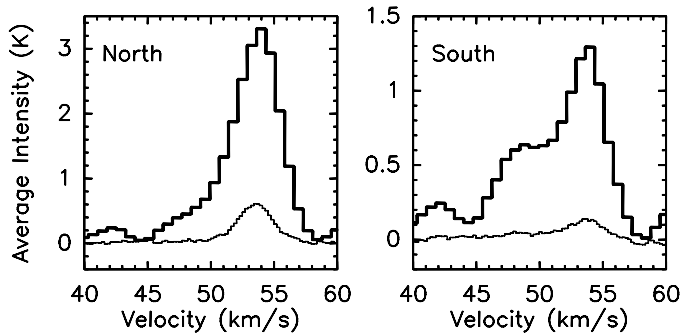,height=1.7in,angle=270, clip=} \hfil\hfil}}
\caption{The 40--60~km~s$^{-1}$ CO spectra from northern
(6.5~arcmin$^2$, {\em left}) and southern (9~arcmin$^2$, {\em
right} ) regions of SNR Kes~75 based on the PMOD observation. The
thick and thin lines are for the $^{12}$CO($J$=1-0) and
$^{13}$CO($J$=1-0) emission, respectively.} \label{f:pmod_54}
\end{figure}

We extracted the \twCO\ and \thCO\ spectra in the velocity range
40--60~km~s$^{-1}$ from two selected regions (as defined in
Fig.~\ref{f:close-up}), which basically correspond to the
X-ray/IR/radio--bright southern half and faint northern half,
respectively. The $^{12}$CO and $^{13}$CO spectra (shown in
Fig.~\ref{f:pmod_54}) in the both regions display a Gaussian
profile around $V_{\rm LSR}\sim$54~km~s$^{-1}$. There is a broad
left wing (45--51~km~s$^{-1}$) in the \twCO\ line profiles for
both the northern and southern regions (as already seen in the
spectrum for the entire remnant field, Fig.~\ref{f:srao_sp}). Similar wing
broadening is not seen in the \thCO\ lines. The
$\sim49$--58~km~s$^{-1}$ \twCO\ and \thCO\ Gaussian lines of the
northern region are stronger than those of the southern region,
indicating that the CO emission come majorly from the northern
region, consistent with the $^{12}$CO channel maps
(Fig.~\ref{f:srao_im1}). However, the broadened left wing is not
as strong in the northern region as in the southern one except in
the interval 49--51~km~s$^{-1}$. For the defined southern region,
we obtain the mean integrated intensity spanning the line profile
(45--58~km~s$^{-1}$) $\sim$8.2~K~km~s$^{-1}$ and the corresponding
molecular column density and gas mass as listed in Table~2. The
contribution of the Gaussian component centered at 54~km~s$^{-1}$
and the broadened part (with the Gaussian component subtracted)
are also estimated, respectively, as given in Table~2.

\begin{figure}[tbh!]
\centerline{ {\hfil\hfil \psfig{figure=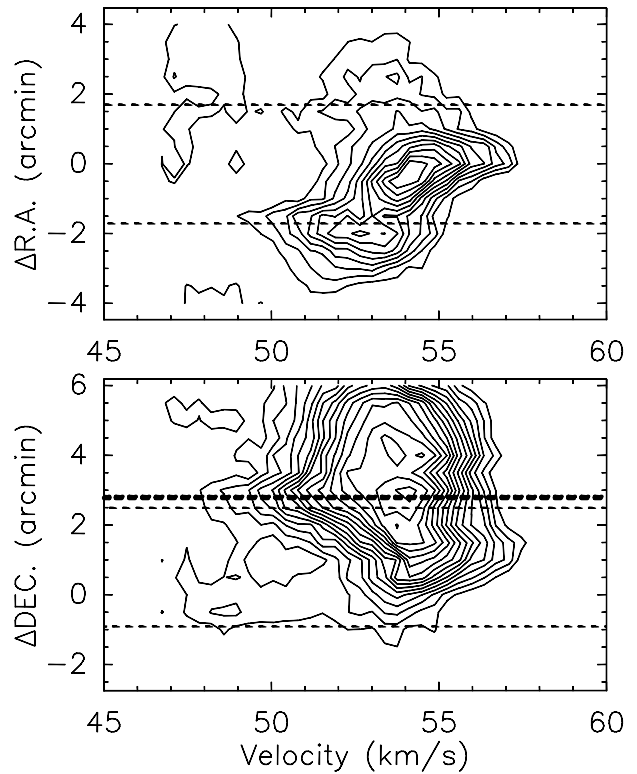,height=4.in,angle=0,
clip=} \hfil\hfil}} \caption{Position-velocity diagrams along E-W
cut (upper panel) and N-S cut (lower panel) [centered at
($\RA{18}{46}{25}.0,\Dec{-02}{58}{48}$)] based on the PMOD
observation. Contour levels start from 1.0~K with in steps of
0.4~K. The dashed lines mark spatial the region of radius $1.7'$
from the pulsar and the thick line marks 2.0$'$ (see
\S\ref{physics}).} \label{f:pv}
\end{figure}

To better understand the velocity structure of the second \twCO\
component, we made  the position-velocity (PV) diagrams
(Fig.~\ref{f:pv}) along the east-west (E-W) and north-south
(N-S) cuts centered at ($\RA{18}{46}{25}.0,\Dec{-02}{58}{48}$) (as
labeled in Fig.~\ref{f:close-up}). In Figure~\ref{f:pv}, it is
clearly shown that the main body of the MC is located in the north
of the Kes~75 area and the maximums of the \twCO\ line broadening
appear near the edge (at radius about 1.7--$2'$ from the pulsar) of
the remnant.

\begin{figure}[tbh!]
\centerline{\mbox{\hspace{25mm}} {\hfil\hfil \psfig{figure=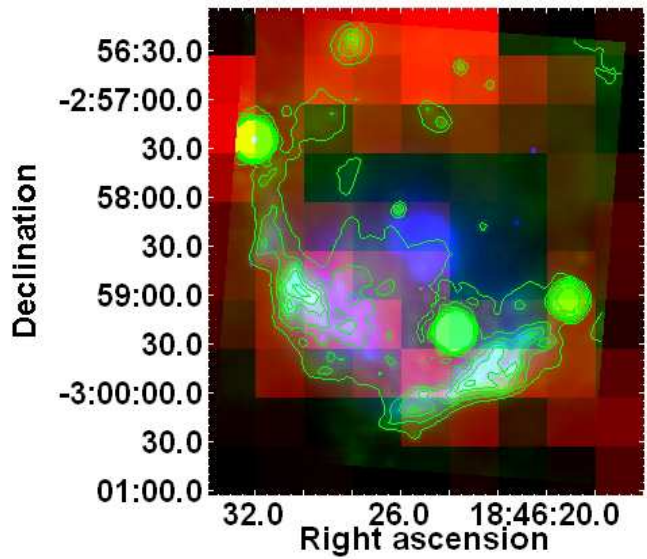,height=3.2in,angle=0,
clip=} \hfil\hfil}} \caption{Tri-color image of SNR Kes~75. The
PMOD $^{12}$CO intensity map ($>2.2$~K~km~s$^{-1}$ or $11\sigma$)
in the interval 45--51~km~s$^{-1}$ is in {\em red}, the {\sl
Spitzer} 24$\mu$m mid-IR emission in {\em green} (including green
contours), and the {\sl Chandra} 1.0--7.0~keV
X-rays (adaptively smoothed with
the CIAO program {\em csmooth} to achieve a S/N ratio of 3) in
{\em blue}. Each of the intensity maps is in logarithmic scales.}
\label{f:tri}
\end{figure}

The close-up \twCO\ channel maps and the integrated map in the
blue-wing interval 45--51~km~s$^{-1}$ are presented in
Figure~\ref{f:close-up}, in which the 6-cm radio continuum contours are
superposed on the integrated map. In Figure~\ref{f:tri} is shown a
tri-color image of the Kes~75 area, with the $^{12}$CO intensity in
the interval 45--51~km~s$^{-1}$ in {\em red}, the {\sl Spitzer}
24$\mu$m mid-IR intensity (together with the contours) in {\em
green}, and the {\sl Chandra} X-ray emission in {\em blue}. A cavity
is revealed to be surrounded by a molecular shell in the line wing
(45--51~km~s$^{-1}$) intensity map, and, surprisingly, the shell in
the south is coincident with the bright rim seen in the X-ray,
mid-IR, and radio emissions. This shell is responsible for the
maximums of the line broadening in the PV diagrams. The bright \twCO\
patch on the northern edge of the SNR may contain the contamination
from the contribution of the main body of the $\sim54$~km~s$^{-1}$
MC at its Gaussian wing 49--$51$~km~s$^{-1}$ ({\em left panel} of
Fig.\ref{f:pmod_54}), which has been seen to be located in the north
(Fig.\ref{f:srao_im1} and Fig.\ref{f:pv}).

We did not find any CO feature between 60~km~s$^{-1}$ and
110~km~s$^{-1}$ (i.e.\ for the third molecular component)
positionally coincident with the southern X-ray/mid-IR/radio--bright
rim. The clouds in this velocity range are estimated to have a
column density $N({\rm H}_2)\sim2.4\E{22}$~cm$^{-2}$ in the northern
region (as defined in Fig.\ref{f:close-up}) and
$\sim1.9\E{22}$~cm$^{-2}$ in the southern region
(Fig.\ref{f:close-up}).

\begin{center}
\begin{deluxetable}{cccc}
\tablecaption{{\sl Chandra} X-ray Spectral Fits with 90\% Confidence
Ranges}
\tablehead{\colhead{Parameter}
          &\colhead{SE Clump}
          &\colhead{SW Clump}
          &\colhead{Northern Diffuse}}
\startdata $N_H$(10$^{22}$cm$^{-2}$) & $3.03^{+0.05}_{-0.13}$ &
$2.72^{+0.20}_{-0.17}$ & $3.81^{+0.47}_{-0.32}$\\
$kT_x$(keV) & $0.65^{+0.03}_{-0.02}$ &
$0.88^{+0.30}_{-0.21}$ \\
$Mg$ &  $1.26^{+0.18}_{-0.14}$  &  $1.04^{+0.24}_{-0.16}$ \\
$Si$ &  $1.79^{+0.28}_{-0.26}$  &  $2.88^{+1.22}_{-0.83}$ \\
$S $ &  $2.75^{+0.37}_{-0.40}$  &  $7.16^{+4.57}_{-2.36}$ \\
$\tau$(10$^{11}$s~cm$^{-3}$) &$4.12^{+1.23}_{-1.10}$ &
$0.35^{+0.10}_{-0.07}$ \\
$norm$(10$^{-2}$) &$1.36^{+0.35}_{-0.44}$ &
$0.18^{+0.19}_{-0.05}$ \\
$PhoIndex$ &$2.76^{+0.29}_{-0.14}$ &
$2.80^{+0.07}_{-0.07}$ & $2.75^{+0.29}_{-0.20}$\\
$norm$ (10$^{-3})$&$1.92^{+0.78}_{-0.51}$ &
$1.54^{+0.18}_{-0.14}$ & $0.56^{+0.28}_{-0.14}$\\
$\chi_r^{2}$/dof &1.19/277 &
1.14/244 & 0.86/81\\
\hline \hline $N_H$(10$^{22}$cm$^{-2}$) & $3.23^{+0.02}_{-0.05}$ &
$3.55^{+0.08}_{-0.11}$ \\
$kT_x$(keV) & $0.60^{+0.04}_{-0.03}$ &
$0.32^{+0.21}_{-0.06}$ \\
$\tau$(10$^{12}$s~cm$^{-3}$) & $>2.8$ &
$ >1.0 $ \\
$norm$(10$^{-2}$) &$1.84^{+0.28}_{-0.27}$ &
$3.97^{+2.81}_{-2.73}$ \\
$kT_x$(keV) & $2.82^{+0.33}_{-0.27}$ &
$2.21^{+0.20}_{-0.20}$ \\
$Mg$ & $1.36^{+0.27}_{-0.26}$  & $<0.42$ \\
$Si$ & $1.67^{+0.128}_{-0.16}$  & $1.11^{+0.18}_{-0.16}$  \\
$S$ &  $2.39^{+0.23}_{-0.34}$  & $1.72^{+0.23}_{-0.25}$  \\
$\tau$(10$^{10}$s~cm$^{-3}$) &$3.53^{+0.36}_{-0.44}$ &
$3.41^{+0.56}_{-0.37}$ \\
$norm$(10$^{-3}$) &$2.78^{+0.47}_{-0.46}$ &
$3.35^{+0.57}_{-0.58}$ \\
$\chi_r^{2}$/dof &1.15/277 &
1.26/244 \\
\enddata
\end{deluxetable}
\end{center}

\subsection{X-ray Spectral Properties of Kes~75}\label{xray}

\begin{figure}[tbh!]
\centerline{ {\hfil\hfil
\psfig{figure=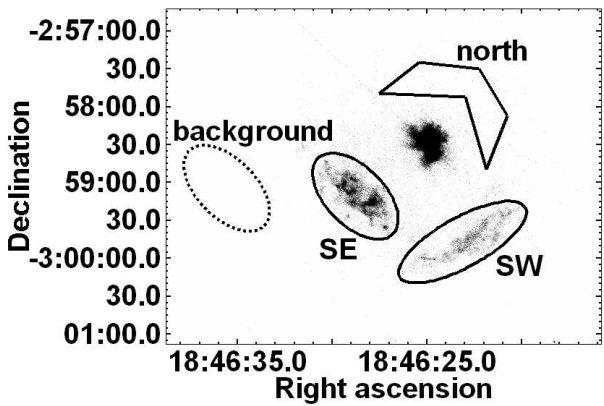,height=2.2in,angle=0, clip=} \hfil\hfil}}
\caption{The {\sl Chandra} X-ray raw image of SNR~Kes~75, labeled
with the defined regions for the X-ray spectrum extraction. The
point sources (except the central pulsar) detected with a wavelet
source-detection algorithm have been removed.} \label{f:xray}
\end{figure}

To investigate the positional relation of the X-ray properties of
SNR~Kes~75 with the CO intensity distribution, we extracted three
X-ray spectra of the southern clumpy shell and the faint northern
diffuse emission (Fig.~\ref{f:xray}) from the reduced {\sl Chandra}
data using the CIAO3.4 script {\sl specextract}. The background
spectrum was extracted from a southeastern region outside the shell.
The three on-source spectra are adaptively regrouped to achieve a
background-subtracted signal-to-noise ratio (S/N) of 5 per bin. The
previous {\sl Chandra} X-ray studies by Helfand et al.\ (2003) and
Morton et al.\ (2007) both used a uniform absorbing hydrogen column
density $N_H$, fixed to that for the central PWN, for the entire
remnant. Here we will allow $N_H$ to vary for each region in our
spectral fit, in which the XSPEC11.3.2 software package with the
Morrison \& McCammon (1983) interstellar absorption is used.

\begin{figure}[tbh!]
\centerline{ {\hfil\hfil \psfig{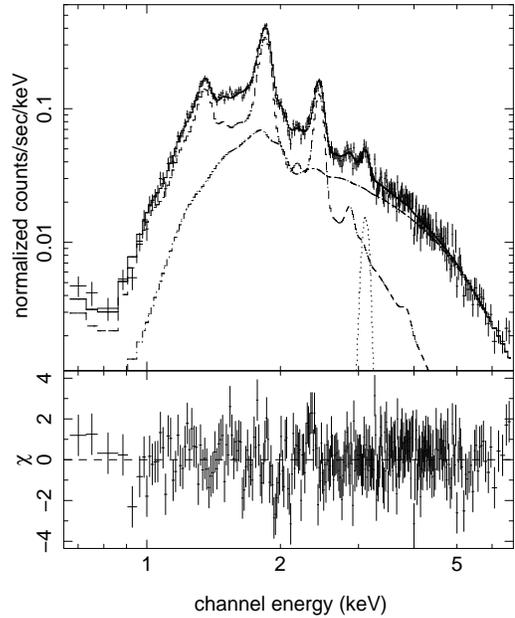} \hfil\hfil}} \caption{The {\sl Chandra}-ACIS X-ray spectrum
of the southeastern clump of Kes~75 (taken from a 183~ks data),
fitted with the {\em vnei} (dashed line) plus {\em power-law}
(dashed-dotted line) model. The dotted line around 3.12~keV is a
Gaussian accounting for the Ar XVII line.} \label{f:se}
\end{figure}

\begin{figure}[tbh!]
\centerline{ {\hfil\hfil
\psfig{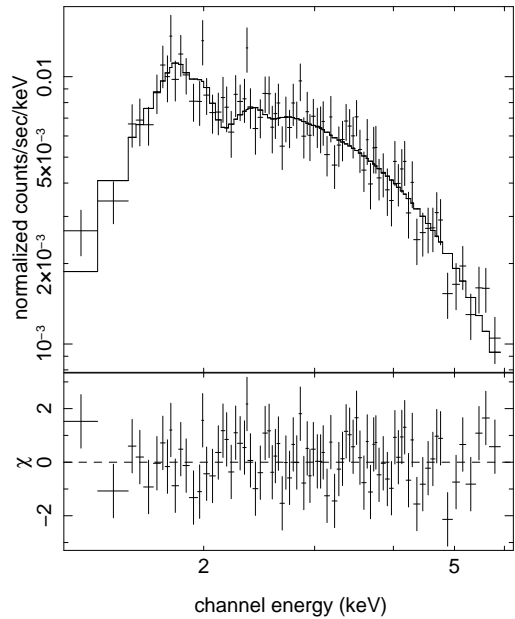} \hfil\hfil}}
\caption{The {\sl Chandra}-ACIS X-ray spectrum of the northern
region of Kes~75 (taken from a 183~ks data), fitted with a
non-thermal {\em power-law} model.} \label{f:n}
\end{figure}

The spectra of the two clumps along the southern shell contain
distinct He$\alpha$ lines of Mg, Si, S, and Ar (see Fig.\ref{f:se}
for the southeastern clump), as have been seen in the previous
studies. However, the spectrum of the northern region appears to
be featureless (Fig.\ref{f:n}). The northern diffuse emission may
be the dust-scattered light from the PWN; actually, Helfand et
al.\ (2003) have pointed out that the effect of the
dust-scattering halo of the Kes~75 PWN should be taken into
account. Recently, Seward et al.\ (2006) clearly detected a
dust-scattered X-ray halo around the Crab comprising 5\% of the
total strength, with surface brightness measured out to a radial
distance of $18'$. Therefore we fit the featureless spectrum of
the northern region with a {\em power-law} model and, following
Helfand et al.\ (2003), fit the spectra of the southern clumpy
shell with a {\em vnei} + {\em power-law} model. As an
alternative, we also fit the two spectra of the southern shell
with a double-thermal-component model (accounting for the forward-
and reverse-shocked gas), as Morton et al.\ (2007) did. The
parameters of the fitting (with the 90\% confidence ranges)
are listed in Table 3.

The both models of our spectral fit generate a lower intervening
hydrogen column density for the southern regions
($\le3.6\E{22}$~cm$^{-2}$) than that used in the previous studies
(fixed to the value for the PWN $\sim4.0\E{22}$~cm$^{-2}$). The
northern diffuse emission has an intervening hydrogen column
$N_{\rm H}\sim
3.5$--$4.3\E{22}$~cm$^{-2}$, which seems
higher than that for the southern regions.
The column estimate for the southern regions from the
{\em vnei}+{\em powerlaw} model, 2.6--$3.1\E{22}$~cm$^{-2}$,
is clearly lower than the column for the northern region.
It is notable that the
difference of $N_H$ between the northern region of the remnant
(including the central PWN) and the southern regions in the case of
using the {\em vnei}+{\em powerlaw} model is around
$\sim1\E{22}$~cm$^{-2}$ (the differnce between the northern and
southwestern regions is $1.1^{+0.5}_{-0.4}\E{22}$~cm$^{-2}$
and that between the northern and southeastern regions is
$0.8^{+0.5}_{-0.4}\E{22}$~cm$^{-2}$).
The both models produce elevated Si and S
abundances. Using the volume emission measure, $fn_{e}n_{H}V$,
derived from the normalization of the XSPEC model, we obtain an
estimate for the hot-gas hydrogen-atom density
$n_H\sim5.4/1.4f^{-1/2}\du^{-1/2}$~cm$^{-3}$ for the
southeastern/southwestern regions, where $n_e$ is the electron
density, $V$ is the volume of
the emitting region, and $f$ is the volume filling factor of the hot gas
($0<f<1$). Here we have assumed $n_{e}\approx1.2n_{H}$ and oblate
spheroids ($24''\times42''\times42''$ and
$19''\times57''\times57''$, respectively) for the defined
southeastern and southwestern elliptical regions
(Fig.~\ref{f:xray}). The mass of the southeastern/southwestern X-ray
bright region is $\sim3.4/1.7f^{1/2}\du^{5/2}M_{\odot}$. The
ionization timescales of the southeastern/southwestern
X-ray--brighten regions in the {\em vnei}+{\em powerlaw} model thus
imply ages of $\sim2.0/0.6f^{1/2}\du^{1/2}$~kyr, which are
comparable with the spin-down timescale 723~yr (Gotthelf et al.\
2000), considering factor $f$ is uncertain.

\section{DISCUSSION}

\subsection{Association of SNR Kes~75 with the $V_{\rm LSR}
 \sim$54~km~s$^{-1}$ MC and the Distance}\label{distance}

From the above results of CO observations, we have obtained
evidences for the association between SNR Kes~75 and the $V_{\rm LSR}
 \sim$54~km~s$^{-1}$ MC. First, as seen in the \twCO\ spectra
(Figs.~\ref{f:srao_sp} \&~\ref{f:pmod_54}), the broadened blue
wing of the second (54~km~s$^{-1}$) MC component indicates that
the MC contains a part of gas that suffers a perturbation toward the
observer. Similar one-sided broadened velocity-profiles are seen
from a piece of MC in SNR~CTB~109 (Fig.~2 in Sasaki et al.\ 2006)
and the molecular arcs in SNR~Kes~69 (Paper~I).
Secondly, the PV diagrams (Fig.~\ref{f:pv}) suggest that this
blue-wing broadening takes place chiefly near the edge of the Kes~75
remnant. These characteristics hint that the 54~km~s$^{-1}$ MC
component may be perturbed by some interaction
related with SNR~Kes~75.
Thirdly, the image of this broadening part is characterized
by a shell-like structure surrounding a cavity, the southern part
of which is coincident with the partial SNR shell seen in X-ray, mid-IR, and
radio continuum (Figs.~\ref{f:close-up} and \ref{f:tri}).
This coincidence not only, again, demonstrates a consistent
spatial position with that of the SNR, but also implies
a shock interaction between the SNR and the molecular material
(detailed in \S\ref{physics}).

The establishment of the association of the $V_{\rm
LSR}\sim$54~km~s$^{-1}$ MC with SNR Kes~75 can provide an
independent estimate for the kinematic distance to the SNR.
Corresponding to the line center $V_{\rm LSR}\approx$54~km~s$^{-1}$
of the unperturbed Gaussian profile (\S\ref{54_MC}), there are two
candidate kinematic distances to the MC along the LOS in the
direction of Galactic longitude $l=29.7^{\circ}$. Using the
rotation curve in Clements (1985), the two candidate
distances are 3.3 and 10.6~kpc,
which fall in the near and far side of the
Sagittarius arm, respectively. Here we refer to the spiral model
in Taylor \& Cordes (1993; Fig.~1 therein) and adopt the circular
rotation model with $R_0=8.0$~kpc (Reid 1993) and the constant
circular velocity $V_0=220$~km~s$^{-1}$.  The tangent point in
this direction is at $\sim7.0$~kpc at the LSR velocity
$\sim103$~km~s$^{-1}$. The HI observation reveals prominent
absorption features at 68, 81, and 95~km~s$^{-1}$ (e.g., LT08).
Because these absorption features are seen at velocities greater than the
systemic velocity, 54~km~s$^{-1}$, of the MC associated with
Kes~75, the SNR is located at the far side of the Sagittarius arm.
This location is consistent with the lower limit, $\sim5.1$~kpc,
placed on the distance to this remnant by LT08 based on HI
absorption.
Thus, we determine the distance to be 10.6~kpc. As a consequence,
the 60--110~km~s$^{-1}$ MCs are all located in front of the
54~km~s$^{-1}$ MC/Kes~75 association.

\begin{figure*}[tbh!]
\centerline{ {\hfil\hfil
\psfig{figure=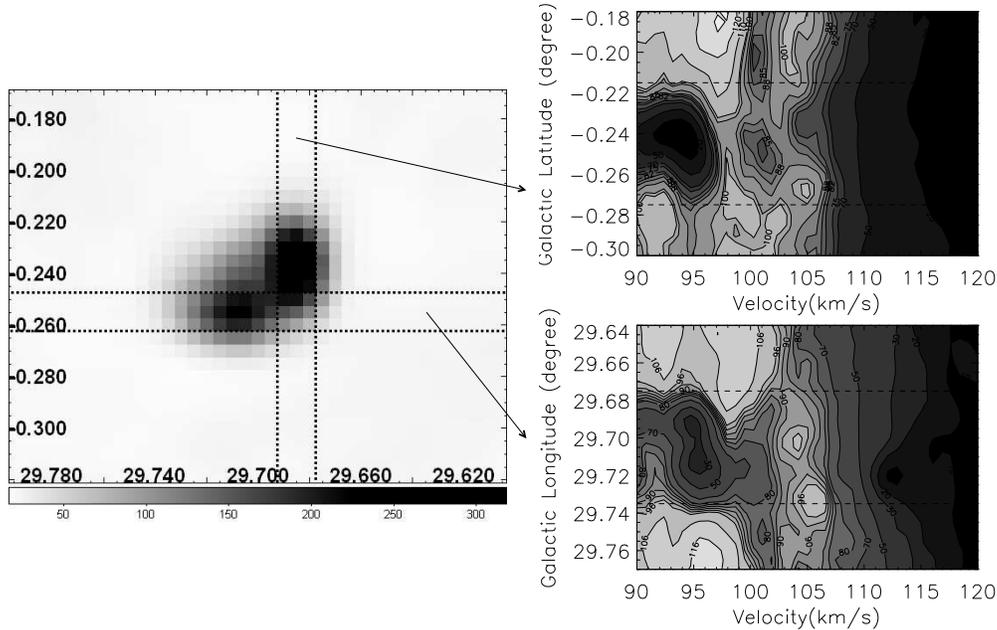,height=3.5in,angle=0, clip=} \hfil\hfil}}
\caption{The 1420 MHz continuum image ({\em left}) and the
position-velocity diagrams of HI emission along the Galactic
latitude ({\em upper-right}) and longitude ({\em lower-right})
cuts. In the two PV dragrams, HI emission is shown with contours
at 20, 50, 70, 75, 82, 85, 88, 92, 98, 100, 110, and 120 and 10,
20, 30, 50, 70, 80, 84, 87, 90, 93, 96, 106, and 116~K,
respectively, with the dark greyscale levels representing low
brightness temperature. The dashed lines mark the spatial extent
of the SNR. The rms noise of the PV diagrams is 1.15~K.}
\label{f:HI-PV}
\end{figure*}

Our distance estimate is larger than the previous estimate of the
upper limit, $\sim7.5$~kpc, set by LT08, which suggests no HI absorption
by the CO-line emitting cloud around 102~km~s$^{-1}$ in the
direction of Kes~75. The HI spectra in LT08 were extracted from
two small regions which are bright in the radio continuum.
To inspect the absorption in the larger radio-bright field, we
made two-dimensional analysis of HI spectra along the Galactic
latitude and longitude cuts near Kes~75, which are 3-pixels wide
and cover most part of the radio-bright region (as shown in
Fig.~\ref{f:HI-PV}). We see in the produced HI PV diagrams that,
above 95~km~s$^{-1}$ at which the prominent HI absorption appears,
there is a clear absorption feature at 100--103~km~s$^{-1}$.
Because of this feature, the HI absorption at
$\sim102$~km~s$^{-1}$, although not strong, cannot be ignored, and
this gas (at a distance of either 6.4~kpc or 7.5~kpc) is thus
allowed to be located in the foreground of the SNR.

\subsection{SNR physics of Kes~75}\label{physics}

The discovery of the MC-SNR association and the updated distance
to Kes~75 are helpful in understanding the remnant's physical properties.

The SNR has a semicircular morphology in {\sl Chandra} X-ray,
24$\mu$m mid-IR, and 1.4~GHz radio wavelengths. It might naturally
have been suspected that the SNR interacts from the southern side, and
slowed by, the huge MC. A similar scenario was suggested for SNR~CTB~109,
in which the SNR interacts with the western massive MC
(Tatematsu et al.\ 1987; Wang et al.\ 1992; Rho \& Petre 1997;
Sasaki et al.\ 2004).
In this case, for the young (\S\ref{xray})
SNR~Kes75, because of the drastic shock deceleration and hence the
magnetic field compression and amplification, the radio emission
might have been expected to be strong along the interface between
the SNR and the MC. Also, because of the conversion of kinetic
energy to the thermal energy, the X-rays might have been expected
to be bright near the interface.
In the CTB~109, the SNR is so evolved ($\sim1\E{4}$~yr) that
the X-ray and radio emission have become faint along the border.
The enhanced emissions are actually seen
along the western rim of the middle age ($\sim4$~kyr) SNR~3C391,
where the blast wave has been suggested to hit a dense cloud plane
(Chen et al.\ 2004). In Kes~75, however, the enhancement in either
the radio or the X-ray emission does not appear along the northern edge
of the apparent semicircle (if this edge were supposed as the interface).

\begin{figure}[tbh!]
\centerline{ {\hfil\hfil \psfig{figure=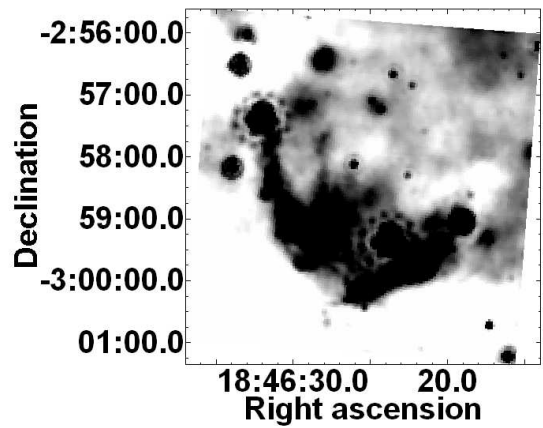,height=2.3in,angle=0,
clip=} \hfil\hfil}} \caption{The grey-scale {\sl Spitzer} 24$\mu$m
mid-IR image in logarithmic scales.} \label{f:ir}
\end{figure}

An alternative possibility is that the SNR
is located behind the MC and the X-ray of the northern half
suffers severe extinction. In the latter case, however, since the
effect of extinction by the cloud is negligible at the 1.4~GHz,
the northern radio emission should be intrinsically faint.
Therefore the SNR might propagate into a northern low-density
region behind the MC, and this scenario thus explain why
the remnant is also faint in the north in X-ray and mid-IR.
It appears that faint
24$\mu$m mid-IR diffuse emission can marginally be discerned in
the northern region (see Fig.\ref{f:ir}), but of course there may
be confusion between the 24$\mu$m emission from the SNR itself
and that from the larger molecular cloud region. In addition, a short,
broken structure seems to be present in the 89~GHz radio image
(Fig.3 in Bock \& Gaensler 2005), coincident with the northern
patch of the bright $^{12}$CO blue-wing emission (i.e., a weak CO-arc
emission probably there, \S\ref{54_MC}), namely, related to the
northern maximum broadening in the CO PV maps (along the {\em thick}
dashed line in Fig.\ref{f:pv}; \S\ref{54_MC}). If this radio
structure is a part of broken shell on the northern rim of the SNR, then
this part of shell and the southern partial shell (in X-ray, IR, and
radio) delineate a roughly circular morphology of the SNR.

In the south, on the other hand, the partial SNR shell seen in
these wavelengths has been shown to be coincident with the
molecular shell in the blue wing of the 54~km~s$^{-1}$ CO emission
(Fig.\ref{f:tri}).
The strong radio continuum emission may be generated
from the SNR shock that is slowed by dense clouds
(Frail \& Mitchell 1998) or the blast wave that propagates
in the intercloud material (Blandford \& Cowie 1982).
The molecular arc moving at a velocity $v_m$ of order 10~km~s$^{-1}$
(implied by the blue-shifted line broadening which reflects
the velocity component along the LOS), as a possibility,
may represent the molecular gas that is perturbed by the slow
transmitted cloud shock after it is struck by the blast wave.
The position of the southern molecular shell may also indicate that
some dense molecular gas may have been engulfed by the blast wave.
The thermal X-ray emission may arise from the heated and evaporated
gas (together with the ejecta) behind the shocked molecular gas,
surrounding the engulfed clumps, or just behind the blast wave.
In this scenario, there can be a crude pressure balance between the
cloud shock and the X-ray emitting hot gas
(Zel'dovich \& Raizer 1967; McKee \& Cowie 1975):
$1.4n_m m_H v_m^2\sim2.3n_H kT_x$,
where $n_m$ is the number density of the hydrogen atoms
ahead of the cloud shock, and $kT_x$ the temperature of the X-ray
emitting gas (see Table 3; here the {\em vnei}+{\em powerlaw} model
is used).
Thereby $n_m\sim4\E{3}(n_H/3\,\mbox{cm}^{-3})(kT_x/0.8\,\mbox{keV})
(v_m/10\,\mbox{km}\,\mbox{s}^{-1})^{-2}\,\mbox{cm}^{-3}$.
This is over an order of magnitude
higher than the mean density of the 54~km~s$^{-1}$ main-body MC
[$n({\rm H}_2)\sim60\du^{-1}\,\mbox{cm}^{-3}$, \S\ref{54_MC}].
The 24$\mu$m IR emission may come from the dust grains heated
by the hot gas and even possibly from the shocked molecular gas
(e.g., OH and H$_2$O).

In the above scenario, the dense molecular shell is pre-existing
before the SNR shock arrives. The molecular shell could not
be the material swept up by the supernova
blast wave all the way to this radius ($r\sim1.8'\sim6\du$~pc).
We adopt the observed total mass $\sim2.7\E{3}\du^2 M_{\odot}$
and the Gaussian-subtracted mass $\sim0.9\E{3}\du^2 M_{\odot}$
(Table~2) as the upper and lower limits of the mass of the southern
molecular shell, respectively.
If it were swept up by the blast wave, the original mean density of
the molecular gas (counting a quarter of sphere of radius $1.8'$)
before it is swept up is $n_0({\rm H}_2)\sim60$--$180\du^{-1}\,
{\rm cm}^{-3}$ (the lower limit of the density is similar
to the mean cloud density, \S\ref{54_MC}).
Based on the Sedov (1959) evolutionary law,
if the blast-wave expansion velocity $v\sim3.7\E{3}\,{\rm km~s}^{-1}$
is adopted from the FWHM of the Si line (Helfand et al.\ 2003),
the age of the remnant would be $\sim0.4r/v\sim640$~yr,
seemingly close to the pulsar's spin-down time
scale 723~yr (Gotthelf et al.\ 2000).
However, the explosion energy
$E=(25/4\xi)\,2.8n_0(\mbox{H}_2)m_Hv^2r^3
\sim1.3\E{54}[n_0(\mbox{H}_2)/100\,\mbox{cm}^{-3}]
(v/3.7\E{3}\,\mbox{km}\,\mbox{s}^{-1})^2 (r/6\,\mbox{pc})^3$~ergs
(where $\xi=2.026$), unreasonably higher than
the canonical number $10^{51}$~ergs.
On the other hand,
the amount of hot gas in the southern clumps (contributing most
to the thermal X-ray emission from the remnant), $\sim5M_{\odot}$
(\S\ref{xray}), is over two order of magnitudes smaller than
that of the southern molecular shell.
This would imply that massive swept-up molecular gas has quickly
reformed after they were dissociated and ionized by the blast shock
(at a velocity higher than $\sim50$~km~s$^{-1}$, Draine \& McKee 1993).
It seems impossible, however, because the timescale of H$_2$ gas
formation, $\geq10^{5.4}$~yr (Koyama \& Inutsuka 2000),
is much higher than the spin-down time (i.e., the remnant's age).

We suggest that the pre-existing molecular shell may be the cooled,
condensed, and clumpy material swept up by the stellar wind of
the progenitor, which was in the vicinity of the large MC,
in supplement to the above scenario of shock hitting the dense
molecular gas.
The presence of the central pulsar evidences that the Kes~75
SNR is the offspring of a core-collapse supernova explosion
of a massive star.
Massive stars may create a cavity with their energetic stellar
winds and ionizing radiation before they explode, and the cavity
wall may be an imprint of the wind modification left on the
nearby environment.
An excellent existing example of the molecular wind-bubble shells
may be that recently discovered coincident with the ring nebula
G79.29+0.46, which surrounds a luminous blue variable star (Rizzo et
al.\ 2008), although it has a smaller radius ($\sim1$~pc) and is
at a higher kinetic temperature.
As a matter of fact, in the case of Kes~75, a cavity is indeed seen
surrounded by a molecular shell (\S\ref{54_MC}; Fig.\ref{f:tri}).
The scenario of the molecular shell of wind bubble
also agrees with the suggestion of a strong Wolf-Rayet
wind prior to the type Ib/c supernova event of Kes~75 (Chevalier
2005).
In Kes~75, there seems
not to be a red-shifted part of shell that expands backward and
thus the wind-bubble may deviate from the spherical
symmetry in the back side.
To catch the essence, some parameters are estimated according to
canonical bubble relation (Castor et al.\ 1975; Weaver et al.\ 1977),
as a crude approximation, as below. The stellar
wind bubble of the Kes~75 SNR progenitor has an age
$\sim5\E{5}(r/6\,{\rm pc})^{5/3}L_{36}^{-1/3}[n({\rm
H}_2)/100\,{\rm cm}^{-3}]^{1/3}$~yr, where $L_{36}$ is the
mechanical luminosity of the stellar wind in units of
$10^{36}\,{\rm ergs}\,{\rm s}^{-1}$. The bubble shell expands at a
velocity of $\sim7(r/6\,{\rm pc})^{-2/3}L_{36}^{1/3}[n({\rm
H}_2)/100\,{\rm cm}^{-3}]^{-1/3}\,{\rm km~s}^{-1}$, a component of
which projected to the LOS may be similar to the blueshift in the
broadened line profile of the disturbed molecular gas.
In addition, the wind-cavity wall scenario could not only naturally
explain why the pre-shock molecular gas hitted by the SNR shock
is over an order of magnitude denser than the main-body MC,
but also avoid the above very high explosion energy by virtue of
the low density (e.g., of order 0.1~cm$^{-3}$ for hydrogen atoms)
in the cavity.

Similar scenario has been suggested to explain the physical
properties of SNR~Kes69, which bears resemblance with Kes~75 in that
the molecular arcs are well coincident with the SNR shells and
move at a velocity of order 10~km~s$^{-1}$ (Paper~I).
In addition to Kes~69, a series of other SNRs have been suggested
to be born in pre-existing stellar-wind cavities,
with the shocks hitting the cavity walls,
such as N132D (Hughes 1987; Chen et al.\ 2003),
W49B (Keohane et al.\ 2007), and Kes~27 (Chen et al.\ 2008).

\begin{figure}[tbh!]
\centerline{ {\hfil\hfil
\psfig{figure=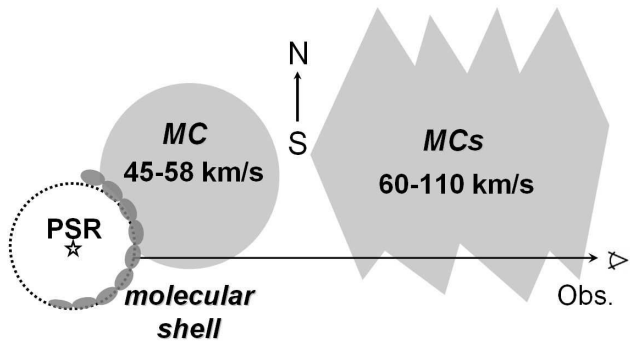,height=1.8in,angle=0, clip=} \hfil\hfil}}
\caption{Diagram showing the positional relation of SNR Kes~75
with the second and third MC complex components.}
\label{f:cartoon}
\end{figure}

In combination of the blue-shifted broadening of the CO line profile,
the SNR is suggested to be located proximately behind the
54~km~s$^{-1}$ MC, and the molecular shell (in the broadened
blue wing) was formed mainly on the near side close to the MC
(i.e.\ toward the observer), as schetched in Figure~\ref{f:cartoon}.
In the north, the remnant
material expands into a tenuous region except for the side close to
the MC, and the northern part of the SNR, together with the
central PWN, is obscured by the MC complexes.

The extra obscuration in the northern area by the MCs could
essentially account for the difference of $N_H$
($\sim1\E{22}$~cm$^{-2}$) between the northern half of the remnant
(including the central PWN) and the southern half (in the case in
which {\em vnei}+{\em powerlaw} model is applied),
as derived from the X-ray spectral analysis (\S\ref{xray}).
This explanation is fairly consistent with the difference of the
molecular column density $\sim1\E{21}$~cm$^{-2}$ between the
northern and southern areas for the second (40--60~km~s$^{-1}$)
molecular component and $\sim5\E{21}$~cm$^{-2}$ for the third
(60--110~km~s$^{-1}$) component (which is located in the
foreground, see \S\ref{distance}).

\subsection{The High-energy Emission of PSR J1846-0258}
McBride et al.\ (2008) find that the most unusual feature of PSR
J1846-0258 is its large efficiency in converting spin-down power
($\dot{E}$) into X- and $\gamma$-ray luminosity ($L_{\rm 0.5-10
keV}$ and $L_{\rm 20-100 keV}$, respectively) of the PWN. In their
work, the SNR distance of 19~kpc was used. If 10.6~kpc is adopted
as the distance to the PSR, the 0.5--10~keV flux of the
Kes~75 PWN $4.0\E{-11}\,{\rm ergs~cm}^{-2}\,{\rm s}^{-1}$ (Helfand
et al.\ 2003) corresponds to a luminosity $L_{\rm 0.5-10
keV}\sim5.4\E{35}\,{\rm ergs~s}^{-1}$ and the X-ray efficiency is
$L_{\rm 0.5-10 keV}/\dot{E}\approx 6.4\%$. (The $\gamma$-ray
efficiency is corrected to $L_{\rm 20-100 keV}/\dot{E}\approx
4.8\%$.) The X-ray efficiency is lower than that ($9.1\%$) of
PSR~B0540$-$69 (Kaaret et al.\ 2001) and similar to that ($5.1\%$)
of the Crab [here $L_{\rm 0.2-10keV}\sim3.3\E{37}\,{\rm
ergs~s}^{-1}$ (Kaaret et al.\ 2001) is converted to $L_{\rm
0.5-10keV}\sim2.4\E{37}\,{\rm ergs~s}^{-1}$ for the Crab Nebula
with the X-ray photon index 2.1 (Willingale et al.\ 2001)], and
thus the efficiency seems to be not peculiar among the young
pulsars.

The High Energy Stereoscopic System (H.E.S.S.) collaboration
(Djannati-Atai et al.\ 2007) discovers the very high energy (VHE)
emission from the sky region of SNR~Kes~75 and regards it as a
point-like source with the point spread function (PSF) $5'$, which
is in a quite good agreement with the position of the PSR
J1846-0258. It is noted that $L_{x}/L{\gamma}\sim$ 10 for the case
of Kes~75, only one-third of SNR~G21.5$-$0.9 and one-twelfth of
the Crab Nebula. This seems to imply a $\gamma$-ray excess in
Kes~75 comparing to other plerions. Since the H.E.S.S.\ VHE
observation cannot resolved the PWN from other structures of
Kes~75, whose apparent size $4'$ is smaller than the PSF, the
$\gamma$ excess cannot be excluded to arise from other sites than
the PWN. The discovery of the association of Kes~75 with the MC
provides a possible interpretation for this excess, because a part
of the $\gamma$-rays may come from the decay of neutral pions
produced by proton interactions between the SNR shock and the MC.

\section{CONCLUSIONS}
Millimeter observations of the molecular gas toward SNR~Kes~75
have been made in SRAO and PMOD. By the spectroscopic analysis of
the \twCO~(J=1-0) emission, in combination with the X-ray, 24$\mu$m
mid-IR, and radio continuum data, we conclude the main results as
follows.
\begin{enumerate}

\item There are several CO emission peaks in a broad LSR velocity
range from 3 to 110~km~s$^{-1}$. The $\sim 83$--96~km~s$^{-1}$ MCs
overlap a large north-western region of the remnant and is
inferred to be located in the foreground of the SNR. The
45--58~km~s$^{-1}$ molecular gas shows a blue-shifted broadening
in the \twCO\ line profile and the intensity centroid of the gas
sits in the northern area of the SNR. The position-velocity map
displays maximums of line broadening near the edge of the remnant.

\item In the integrated intensity map of the broadened blue wing
at 45--51~km~s$^{-1}$, a cavity surrounded by a molecular shell is
unveiled, which is coincident with the SNR, and the shell in the
south follows the bright partial shell seen in X-rays, mid-IR, and
radio continuum.

\item All the observational features of the broadened blue wing
provide effective evidences that Kes~75 is associated with the
adjacent 10$^{4}M_{\odot}$ 54~km~s$^{-1}$ MC.

\item The establishment of the association between Kes~75 and the
54~km~s$^{-1}$ MC results in a determination of the kinematic
distance at $\sim10.6\kpc$ to the remnant, which agrees with a
location at the far side of the Sagittarius arm.

\item The X-ray, mid-IR, and radio shell-like morphology in the
south can be accounted for by the SNR shock striking a
pre-existing molecular shell. This shell may represent the debris
of the cooled, condensed, clumpy shell of the progenitor's
wind-bubble proximately behind the 54~km~s$^{-1}$ cloud.

\item The discovery of the association of Kes~75 with MC can
provide a qualitative explanation for the $\gamma$-ray excess of
the remnant.\\

\end{enumerate}

\begin{acknowledgements}
The authors acknowledge the staff members of the Seoul Radio
Astronomy Observatory and the Qinghai Radio Observing Station at
Delingha for their support in observation. Y.S.\ specially thanks
Jae-Joon Lee for help on CO observation. Wen-Wu Tian and Jun-Zhi
Wang are also thanked for discussion on the distance estimates.
Y.C.\ thanks Q.\ Daniel Wang for discussion and critical reading
of the manuscript. Y.C.\ acknowledges support from NSFC grants
10725312, 10673003, and 10221001 and the 973 Program grant 2009CB824800.
We acknowledge the use of the VGPS data;
the National Radio Astronomy
Observatory is a facility of the National Science Foundation
operated under cooperative agreement by Associated Universities,
Inc.
\end{acknowledgements}



\end{document}